\documentclass[epj]{webofc}
\usepackage[varg]{txfonts}   
\usepackage{lineno}
\begin{document}
%
%
\title{The Crab pulsar at VHE}
%
%

\author{Roberta Zanin\inst{1}\fnsep\thanks{\email{Roberta.Zanin@mpi-hd.mpg.de}}
}

\institute{Max-Planck-Institut fur Kernphysik, P.O. Box 103980, D 69029 Heidelberg, Germany}

\abstract{%

The last six years have witnessed major revisions of our knowledge about the Crab Pulsar. The consensus scenario for the origin of
the high-energy pulsed emission has been challenged with the discovery of a very-high-energy power law tail
extending up to $\sim$400 GeV, above the expected spectral cut off at a few GeV. Now, new measurements
obtained by the MAGIC collaboration extend the energy
spectrum of the Crab Pulsar even further, on the TeV regime. Above $\sim$400 GeV the pulsed emission comes
mainly from the interpulse, which becomes more prominent with energy due to a harder spectral index.
These findings require $\gamma$-ray production via inverse Compton scattering close to or beyond the light
cylinder radius by an underlying particle population with Lorentz factors greater than $5 \times 10^6$. We will
present those new results and discuss the implications in our current knowledge concerning pulsar
environments.
}
\maketitle
\section{Introduction}
\label{intro}

Pulsars are rapidly rotating neutron stars (NS) which
form highly magnetized magnetospheres, usually described as
magnetic dipoles.  
The magnetosphere is filled with charged particles, mainly electrons
and positrons, extracted from the stellar surface by the induced
electric fields. The original model for pulsar magnetospheres foresees
that these charged particles satisfy the ideal
magnetohydrodynamic (MHD) ($E \cdot B$=0) and the free-force (FF,
$\rho E+J \times B$=0) conditions \cite{Goldreich:1969}. 
In this view, the magnetic field lines cannot remained
closed beyond the light cylinder (LC, with a radius R$_{LC}$), where the angular velocity of the NS
is equal to the speed of light. Along the open field lines the plasma
flows along asymptotically monopole field lines \cite{Michel:1973}
forming the \emph{pulsar wind region}. 
Nowadays it is commonly accepted that the ideal MHD and FF conditions
cannot be fulfilled everywhere inside the magnetosphere. In the past years
regions where the electric field is not totally screened by the plasma
were invoked as possible sites of particle accelerations. 
They are referred as to \emph{gaps}. Recently current
sheets are taking over though. They are current carrying surfaces
where particles can be accelerated via relativistic reconnection and
radiate synchrotron emission from optical to $\gamma$-ray wavelengths. 
In particular, an equatorial current sheet was proposed
by Contopoulos to ensure the closure of the large-scale currents
flowing from the star to infinity and back \cite{Contopoulos:1999},
but only recent particle-in-cell (PIC) simulations of plasma-filled
magnetospheres have provided the evidence for its
existence, as well as its possible extension down to the NS surface,
separating the open and closed field lines (this names them
\emph{separatrix})
\cite{Spitkovsky:2006,Kalapotharakos:2009,Tchekhovskoy:2013,Cerutti:2015}. 
The equatorial current sheet separates the field lines that originate from the
two pulsar magnetic poles. When the rotation and magnetic axes are not
aligned (oblique pulsars), this current sheet develops corrugation
whose amplitude increases linearly with the distance from the
star. The equatorial plan is then divided into stripes of magnetic
field of opposite polarities, hence the name \emph{stripped wind}
\cite{Coroniti:1990,Bogovalov:1999,Kirk:2002}. \\
In this new paradigma of the pulsar magnetosphere, two classes of
pulsars are defined depending on the 
particle density inside the magnetosphere: the high-multiplicity
pulsars where pair production takes place within the magnetosphere,
that are the young pulsars with high stellar magnetic fields, and the
low-density pulsars where particles are supplied only from the NS
surface and are identified with the old pulsars. For the first class
of pulsars the FF solution is a good approximation of their
magnetosphere characterized by the presence of the separatrices and the
equatorial current sheet. They are weakly dissipative
\cite{Cerutti:2015,Contopoulos:2016}. The low-density pulsars exhibit
a charged separated magnetosphere with no separatrices since the
equatorial sheet is electrostatically supported. They show high
dissipation ($>$40\%) beyond the LC
\cite{Cerutti:2015,Contopoulos:2016}.  

Pulsars are observed from radio frequencies up to $\gamma$ rays. 
Radio emission is believed to be produced at 
low-altitudes, close to the stellar magnetic poles, by a a coherent
process either maser amplification or coordinated motion of group of
charges. At higher frequencies pulsed emission is incoherent, most
likely synchrotron radiation produced by relativistic electron-positron pairs. 
At high energies, pulsed emission was long attributed to synchro-curvature 
by particles accelerated in magnetospheric gaps. According to the
location of the gap inside the magnetosphere, $\gamma$-ray pulsar models
were classified into \emph{polar caps}, if close to the magnetic poles,
\cite{Harding:1978}, \emph{outer gaps} if at high altitudes
\cite{Cheng:1986}, or \emph{slot gaps}, if along the last
open field lines \cite{Muslimov:2003}. In particular, 
such a synchro-curvature emission has a maximum energy limited by 
radiation losses or magnetic and $\gamma$-$\gamma$ pair absorption. 
The recent advances on the high-energy
pulsar physics due to discovery of more than 150 $\gamma$-ray pulsars by
\emph{Fermi}/LAT \cite{2PC} suggest that pulsed $\gamma$ rays are produced at
high altitudes favouring the \emph{outer gaps} models. This conclusion
is mainly drawn by two observational facts: 1) The spectral cutoff at
a few GeV is exponential, and often even subexponetial for young
pulsars. The smoother spectral break is interpreted as the overlapping
of emission beams coming from different regions and/or with different
cutoff energies. 2) The majority of the
$\gamma$-ray pulsars present a double-peaked pulsar profile, which in
low-altitude emission models would
require a specific geometric configuration (both viewing and
inclination angles close to 90$^{\circ}$), too special to meet
the observed numbers \cite{Romani:2010}. However, in the context of the new
paradigma of the pulsar magnetosphere, the high-energy emission could
entirely or partially be of synchrotron origin produced in the equatorial current sheet
beyond the light cylinder \cite{Petri:2012,Mochol:2015,Cerutti:2016}.


\section{The Crab pulsar}
\label{crab}
The Crab pulsar is the second most powerful pulsar known so far with a
spin-down luminosity $\dot{E}=3.8\times10^{38}$\,erg/s and the only
pulsar whose age is known with precision. It is, in fact, the leftover
of an historical supernova explosion occurred in 1054\,AD which was
in detail documented by Chinese astronomers. 

Its pulse profile is characterised by two peaks almost aligned at
almost all wavelengths, even though their amplitudes change with
energy \cite{Kuiper:2001,bridge}.  The main peak at low radio frequencies
(0.1--5 GHz), by definition set at phase 0, is usually referred to as P1, whereas
P2 or interpulse is located $\sim$\,0.4$^{\circ}$ away. The emission in
between the two peak is called \emph{bridge}. 

\subsection{Recent results at VHE}

All the 160 $\gamma$-ray pulsars detected by \emph{Fermi}/LAT show a
spectral break at $\sim$few GeV in agreement with the
synchro-curvature scenario \cite{2PC}, but one: the Crab pulsar. The
spectrum of the Crab pulsar between 0.1-100\,GeV  is
well parametrized by a power-law function with a sub-exponential
cutoff (b$<$1). The best-fit value for the photon index is
$\gamma$=1.59$\pm$0.01, for the energy break is (5.09$\pm$0.63)\,GeV and for
the curvature index b=0.43$\pm$0.01 \cite{buehler:2012}.
Recent measurements of the Crab pulsar by imaging atmospheric
Cherenkov telescopes showed a totally unexpected power-law component
emerging above the synchro-curvature cutoff and
extending above 100\,GeV \cite{Aliu:2011,Aleksic:2011,Aleksic:2012}. 
This new spectral component, that is excluding the spectral break at
more than than 6$\sigma$ \cite{Aliu:2011}, is a common feature of
both peaks and the bridge. A joint fit of the \emph{Fermi}/LAT data
above 10\,GeV and the MAGIC ones indicates that P2 is significantly
harder than P1 with a difference between the photon indices of
0.5$\pm$0.1.  Whereas the steeper P1 ($\Gamma$=3.5$\pm$0.1) spectrum
cannot be detected beyond 600\,GeV, P2 becomes the dominant component
above 50\,GeV \cite{Aleksic:2011} and extends up to 1.5\,TeV without
any sign of cutoff \cite{MioPSR:2016}. A lower limit on the spectral
cutoff is estimated at 700\,GeV. On the other hand, the spectrum of
the bridge is as soft as P1 and fades out already at 150\,GeV \cite{bridge}.  

This VHE pulsed emission is very unlikely to be synchro-curvature
radiation: the production of 1\,TeV photon would require, in fact, a
curvature radius of the magnetic field lines one order of magnitude
larger than the usual adopted one. If emission up to few hundreds of
GeV could be also synchrotron radiation from the equatorial
current sheet \cite{Petri:2012,Mochol:2015}, the emission at higher energies
is accredited to be produced via inverse Compton scattering. In this
view several models have been put forward. Some of them are listed in
the following:

\begin{itemize}
\item
inverse Compton scattering of relativistic wind electrons off pulsed
optical/X-ray photons with magnetospheric origin
\cite{Bogovalov:2000,Aharonian:2012}. This model was
proposed to explain the emission up to 400\,GeV and can well reproduce
the pulse profile by assuming an anisotropic pulsar wind. Nevertheless,
it fails in reproducing the spectrum up to 1\,TeV. The production of
these energetic photons would require electron parent population with
a Lorentz factor larger than $5\times10^6$, hence a continuos acceleration
from the LC up to $\sim$100R$_{LC}$ which in turn results in an
overestimation of the GeV flux. 
\item
synchrotron-self-Compton scattering off synchrotron photons produced
in the current sheet by the same population of synchrotron-emitting
electrons. This model reproduces the spectrum up to VHE as a sum of
two distinct components: the synchrotron power-law up to few hundreds
of GeV and the SSC as an extra bump fading out at a few TeV
\cite{Mochol:2015}. However, the model cannot explain the pulse
profile. 
\item
 inverse Compton scattering off either photons 
in particle cascading inside the magnetospheric gaps
\cite{Lyutikov:2012,Aleksic:2011,Hirotani:2015}.
\end{itemize}
So far none of the existing model
can explain simultaneously the spectrum and the pulse profile with its
narrow peaks at VHE. In addition there is no comprehensive theoretical
model reproducing all the observable of the pulsed emission i.e. the broadband
spectrum with its spectral changes across the pulse-phase in X- and
$\gamma$ rays, and the pulse profile with its energy-dependent peak
amplitude and width.

\begin{figure}[ht]
\centering
\includegraphics[scale=0.40]{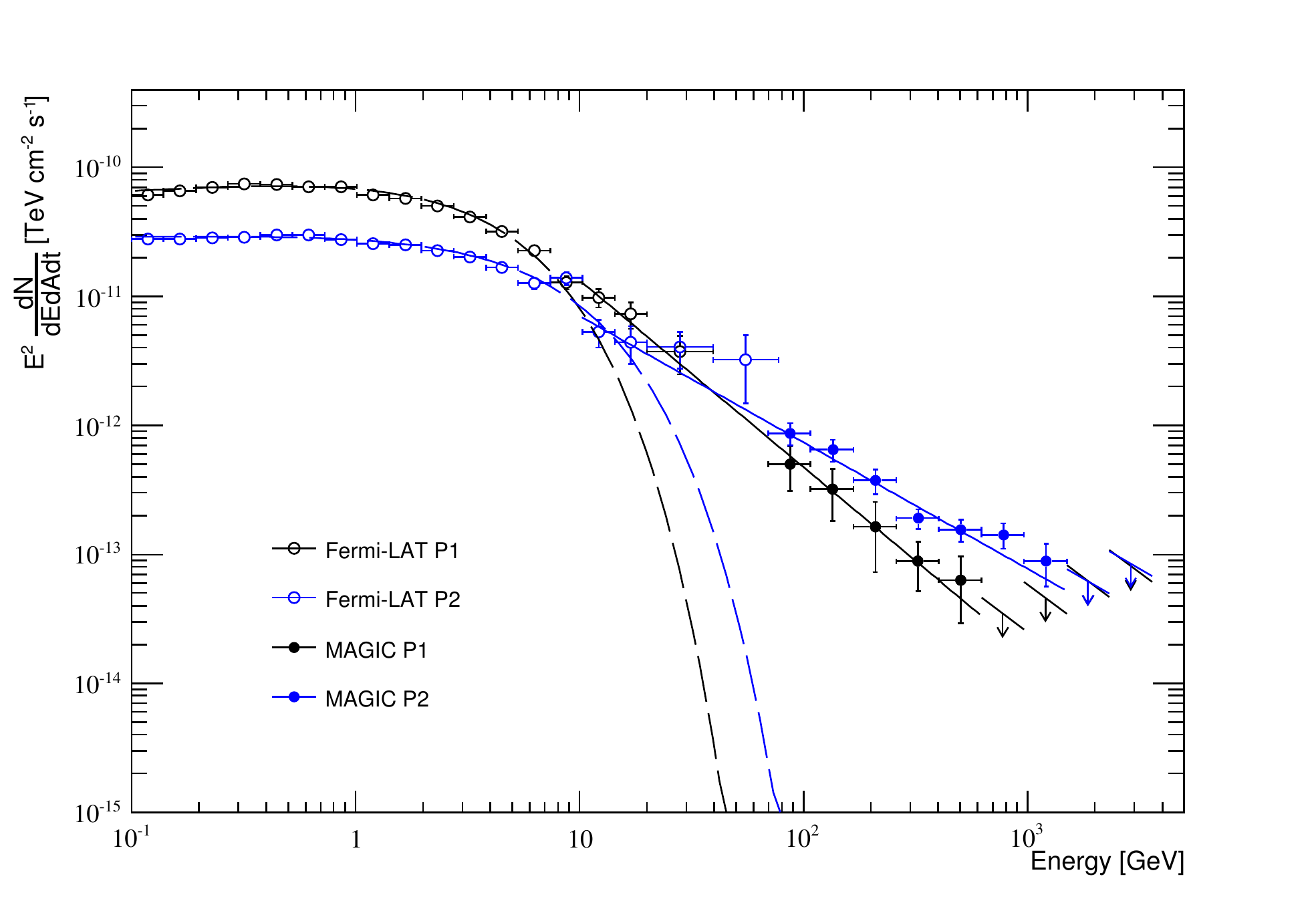}
\caption{Phase-folded SED of the Crab P1 (black circles) and P2 (blue circles) at HE and VHE
(open and filled circles). Taken from \cite{MioPSR:2016}.}
\label{fig-1}       
\end{figure}

 \bibliography{biblio.bib}
%

\end{document}